\documentclass[journal,twocolumn]{IEEEtran}
\usepackage{latexsym}
\usepackage{epsf,epsfig}
\usepackage{citesort}
\usepackage{float}
\usepackage{amsmath,amssymb,graphicx,psfrag}

\newcommand{\bq}{\begin{equation}}
\newcommand{\eq}{\end{equation}}
\newcommand{\bqa}{\begin{eqnarray}}
\newcommand{\eqa}{\end{eqnarray}}
\newcommand{\ben}{\begin{enumerate}}
\newcommand{\een}{\end{enumerate}}
\newcommand{\bl}[1]{\mathbf{#1}}
\newcommand{\bul}[1]{\underline{\bl{#1}}}
\newcommand{\wh}[1]{\widehat{#1}}

\begin{document}
\title{Comments on `Bit Interleaved Coded Modulation'}
\author{Vignesh Sethuraman, Bruce Hajek
\thanks{The first author was supported by the Vodafone-US Foundation Fellowship. The authors are with the University of Illinois at Urbana-Champaign.}}

\maketitle
\footnotetext[1]{DRAFT NOT FOR PUBLIC CIRCULATION: IEEE-IT SUBMISSION }
\begin{abstract}
Caire, Taricco and Biglieri presented a detailed analysis of
bit interleaved coded modulation, a simple and popular technique used to improve system performance, especially in the context of fading channels. 
They derived an upper bound to the probability of error, called the expurgated bound. 
In this correspondence, the proof of the expurgated bound is shown to be flawed. A new upper bound is also derived.                                          
It is not known whether the original expurgated bound is valid for the important special case of square QAM with Gray labeling, but the new bound is very close to, and slightly tighter than, the original bound for a numerical example. 
\end{abstract}
\begin{keywords}
Bit interleaved coded modulation, BICM, expurgated bound, probability of error
\end{keywords}

\section{Introduction} \label{sec.intro}
A comprehensive study of BICM is presented in \cite{Caire98}. There, in addition to an information theoretic analysis of BICM, a detailed analysis of the probability of error is presented. 
In the error analysis of BICM in \cite{Caire98}, various upper bounds and approximations to the probability of error  are derived, notable among which is the expurgated bound. 
In the first half of this paper, counter examples are given for the two theorems in \cite{Caire98} leading to the expurgated bound. Consequently, the validity of the expurgated bound in \cite{Caire98} is questionable. 
The second half of this paper focuses on the important and practical case of square QAM constellations with Gray labeling.
For such cases, an alternate upper bound is presented. Numerical results are given for $16$-QAM and $64$-QAM and a rate-$\frac 1 2$ convolutional code.
For these examples, the new bound is nearly equal to, and slightly tighter than, the expurgated bound of \cite{Caire98}.
The reader is referred to \cite[Sections $2$ and $4$]{Caire98} for notation.

\section{Two counter examples} \label{sec.CounterEg}
\begin{figure}
\begin{center}
\begin{psfrags}
\psfrag{Bad}[c]{${\cal{X}}_1^1$}
\psfrag{Good}[c]{${\cal{X}}_0^1$}
\psfrag{00}[r]{$\bl{x}=00$}
\psfrag{01}[r]{$\bl{z}^{(1)}=01$} 
\psfrag{10}[r]{$\bl{z}^{(2)}=10$}
\psfrag{11}[l]{$11=\bl{z}$}
\epsfig{file=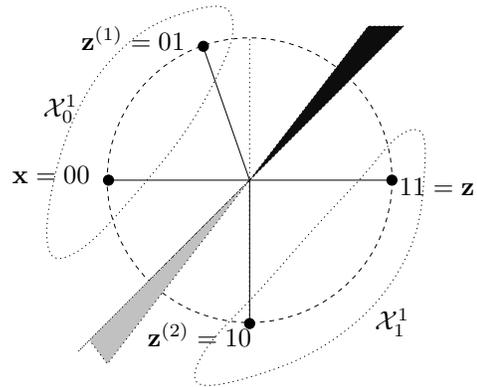,height=2in}
\caption{4PSK -- The shaded regions correspond to the region not covered by the expurgated union bound.} \label{fig:StrangeQpsk}
\end{psfrags}
\end{center}
\end{figure}

{\bf Counter example to \cite[Theorem 1]{Caire98}:}
Consider the constellation in Figure \ref{fig:StrangeQpsk}. 
It is similar to QPSK, except the point $z^{(1)}$ which is on the unit circle but closer to one of its neighbors. 
The labeling $\mu$ is chosen to be Gray labeling. 
For simplicity, assume $d=1$. 
For concreteness, assume that the message to be sent is $b=0$, and that it is to be sent over the first label position, i.e. $S=1$. 
Further, let $U$ be $0$ so that the signal labels are not complemented. 
Then, an element of ${\cal{X}}_0^1 = \{ 00, 01\}$ is transmitted. 
Suppose $\bl{x} = 00$ is transmitted. 
Let $\bl{z}=11$. 
As in \cite{Caire98}, let $\Gamma_{\bul{x},\bul{z}} = \left\{ \bul{y}: \| \bul{y} - \bul{x}\|_2^2 \geq \| \bul{y} - \bul{z}\|_2^2 \right\}$. 
Clearly $\Gamma_{\bl{x},\bl{z}} \subset \Gamma_{\bl{x},\bl{z}^{(1)}} \cup \Gamma_{\bl{x},\bl{z}^{(2)}}$ where $\bl{z}^{(1)}=01$, $\bl{z}^{(2)}=10$. 
So, by \cite[Theorem 1]{Caire98}, $\Gamma_{\bl{x},\bl{z}}$ can be neglected in the union bound.
This leaves only $\Gamma_{\bl{x},\bl{z}^{(2)}}$ in the union bound, but that fails to cover some part of the pairwise error region.
In particular, referring to Figure \ref{fig:StrangeQpsk}, we see that the darkly shaded region is left out of the bound on $P(\mbox{decoder error}|\bl{x}=00)$ though it is part of the pairwise error region. 
Similarly, the lightly shaded region is left out of the bound on  $P(\mbox{decoder error}|\bl{x}=01)$. 
Thus, neglecting the term $P(\bl{x} \rightarrow \bl{z})$ in the union bound alters the inequality. 
This disproves \cite[Theorem 1]{Caire98}. 
The theorem may hold in the presence of  stronger conditions on the constellation such as symmetry. 
We are unable to identify such sufficient conditions though.

The problem with the proof of \cite[Theorem 1]{Caire98} lies in eliminating the sub-region $\{ \Gamma_{\bul{\wh{x}}, \bul{\wh{z}} }  \cap \Gamma_{\bul{\wh{x}}, {\bul{z}}^{(i)}} \}$ where ${\bul{z}}^{(i)} \in {\cal{X}}_{\bul{c}}^{\bul{S}}$. 
Suppose $\bul{y}$ is received. It is reasoned in \cite{Caire98} that, if $|\bul{z}^{(i)} - \bul{y}| < |\bul{x} -\bul{y}|$, then $\bul{y}$ should not be counted in the pairwise error region of $(\bul{x} \rightarrow \bul{z})$  since the decoder either chooses $\bul{z}^{(i)}$ (and thus does not make an error) 
or the decoder makes an error which is 
already	included at least once, in the term $\bul{z}^{(i)} \rightarrow \bul{\wh{z}}$. 

The right hand side of \cite[(30)]{Caire98} can be viewed as an average over the transmitted signal sequence $\bul{x}$, and a sum, or union bound, over $\bul{z}$ in the bad signal subset ${\cal{X}}_{\bul{\hat{c}}}^{\bul{S}}$. 
The decoder error region, for a given $\bul{c}$, $\bul{S}$ and $\bul{U}$, is the same for all transmitted $\bul{x}$, and can be denoted by:
\[
\Gamma_{\bul{c},\bul{\hat{c}}} = \{ \bul{y}: 
		\min_{\bul{x} \in {\cal{X}}_{\bul{c}}^{\bul{S}}} \| \bul{y} - \bul{x} \|^2 \geq
		\min_{\bul{z} \in {\cal{X}}_{\bul{\hat{c}}}^{\bul{S}}} \| \bul{y} - \bul{z} \|^2
	   	\}
\]
When the union bound in \cite[(30)]{Caire98} is expurgated, it is not known whether the inequality remains valid unless all points in $\Gamma_{\bul{c},\bul{\hat{c}}}$ are covered in the expurgated union bound for each transmitted $\bul{x}$. 
So, the region $\Gamma_{\bul{c},\hat{\bul{c}}}$ should be counted $|{\cal{X}}_{\bul{c}}^{\bul{S}}|=2^{d(m-1)}$ times, each time being weighted by a probability measure depending on which $\bul{x} \in {\cal{X}}_{\bul{c}}^{\bul{S}}$ was transmitted.

\begin{figure}
\begin{center}
\epsfig{file=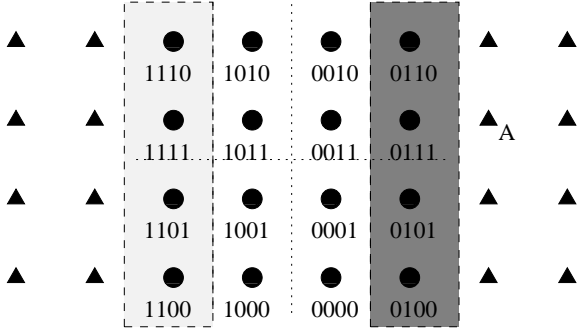,width=3in}
\caption{16QAM -- The shaded regions correspond to the region not covered by the expurgated union bound.} \label{fig:16QAM}
\end{center}
\end{figure}

{\bf Counter example to \cite[Theorem 2]{Caire98}:}
Consider 16QAM constellation with Gray labeling, as given in Figure \ref{fig:16QAM}. 
Let $d=1$, $U=0$ and $b=0$, as before. Let the information be transmitted in $S=2$. So, some $\bl{\wh{x}} \in {\cal{X}}_0^2$ is transmitted. 
Suppose $\bl{\wh{x}}=1011$ is transmitted. Let $\bl{\wh{z}} = 0111$. Setting $\bl{z'}=0010$ and $\bl{z''}=0001$, it can be shown that the conditions of \cite[Theorem 2]{Caire98} are satisfied.
It follows from the theorem  that $P(\bl{\wh{x}} \rightarrow \bl{\wh{z}})$ can be neglected in \cite[(30)]{Caire98} without altering the inequality. It can be similarly concluded that all $\bl{\wh{z}} \in {\cal{X}}_1^2$ except $1111$ can be neglected. 
Referring to Figure \ref{fig:16QAM}, it follows that 
 the darkly shaded region is left out of the bound on $P(\mbox{decoder error}|\bl{\wh{x}}=1011)$, though it is part of the pairwise error region. 
This occurs for all $\bl{\wh{x}}$ of the form $10**$ (where $*$ can be $0$ or $1$). 
Similarly, the lightly shaded region is left out of the bound on  $P(\mbox{decoder error}|\bl{\wh{x}}=00**)$. 
Thus, the inequality does not hold and \cite[Theorem 2]{Caire98} is disproved. 

The extension to fading channels of the above theorem, namely, \cite[\S IV.C Corollary 1]{Caire98}, uses stronger conditions than the above theorem. However, it is easy to see how the above example works as a counter example to the corollary as well. 

The proof that $f_{ex}(d,\mu,{\cal{X}})$ is greater than or equal to $f(d,\mu,{\cal{X}})$ is hence not valid for the case of square QAM signal sets with Gray labeling.
It is to be kept in mind that $f_{ex}(d,\mu,{\cal{X}})$ is presented as an upper bound to $f(d,\mu,{\cal{X}})$ in \cite{Caire98} only for this case - for any other choice of modulation, it is presented as only an approximation. 

\section{Revised expurgated bound} \label{sec.Corrections}
For the case of square QAM signal sets with Gray labeling, a revised expurgated bound, denoted by $f_{ex,new}$, is derived. Two variants of $f_{ex,new}$, denoted by $f_{ex,new}^{I}$ and  $f_{ex,new}^{II}$, are presented. 
First, consider the case when $d=1$. 
Fix a point $\bl{x}$ in the signal set, and a bit position $i$ in its label. 
Since Gray labeling is used, the bit value of the corresponding position remains the same across either the row or the column containing $\bl{x}$. 
So, if $\bl{x} \in {\cal{X}}_b^i$, then either the entire row or the entire column in the signal set belongs to ${\cal{X}}_b^i$. 
This reduces the problem of identifying regions contributing to the probability of error to a single dimensional problem. 
As in any such single dimensional problem, the regions contributing to the probability of error can be covered by choosing two neighbors, one on each side, and constructing a union bound with two PEP terms. 
This is in contrast to the original expurgated bound where only one neighbor is considered.

Suppose $\bl{x} \in {\cal{X}}_b^i$ is transmitted. Then, the original expurgated bound considers only the unique nearest neighbor in ${\cal{X}}_b^i$. 
An easy fix is to consider the nearest neighbors in ${\cal{X}}_{1-b}^i$ on both sides of $\bl{x}$. 
This can be improved further by choosing the neighbors such that the PEP decision boundaries coincide with the actual decision boundaries.
As an illustration, consider the constellation in Figure \ref{fig:16QAM}. As in 
the counter example to \cite[Theorem 2]{Caire98}, suppose $d=1$, $U=0$, $S=2$ and $b=0$.
Some $\bl{x} \in {\cal{X}}_0^2$ is transmitted -- let $1011$ be transmitted. 
Since all the points in the column of $\bl{x}$ belong to ${\cal{X}}_0^2$, the decision boundaries are vertical. The original expurgated bound includes only the unique nearest neighbor in ${\cal{X}}_1^2$, namely $1111$. 
The easy fix is to include the nearest neighbor in ${\cal{X}}_1^2$ on the other side of $\bl{x}$ also - namely $0111$. This variant of $f_{ex,new}$ shall be referred to as $f_{ex,new}^I$. 
Alternatively, the signal set can be extended in a lattice fashion, as shown by triangles in Figure \ref{fig:16QAM}), and  two points in the (extended) signal set can be identified such that the PEP decision boundaries coincide with the actual decision boundaries. For this example, the two points then are $1111$ (as before) and $A$ (instead of $0111$). 
This variant of $f_{ex,new}$ shall be referred to as $f_{ex,new}^{II}$.

The above two methods have their relative merits and demerits. 
While both of the above methods yield upper bounds to $f(1,\mu,{\cal{X}})$ (this follows from a union bound argument), the bounds obtained using the second method will clearly be tighter than the first. 
However, specifying the two points is more straightforward in the first method. 

In either way, there are two points in ${\cal{X}}_{1-b}^i$ corresponding to each  $\bl{x} \in {\cal{X}}_b^i$. 
Let them be denoted by $\bl{z}_1(\bl{x})$ and $\bl{z}_2(\bl{x})$ when referring to $f_{ex,new}$, by $\bl{z}_1^I(\bl{x})$ and $\bl{z}_2^I(\bl{x})$ when referring to the variant $f_{ex,new}^I$, and similarly by  $\bl{z}_1^{II}(\bl{x})$ and $\bl{z}_2^{II}(\bl{x})$ when referring to $f_{ex,new}^{II}$. 
Then, $f_{ex,new}$ can be defined by:
\bqa
f_{ex,new}(1,\mu,{\cal{X}}) = \hspace{1.7in} \nonumber \\
		\frac 1 {m 2^m} \sum_{S,U} \sum_{\bl{x} \in {\cal{X}}_c^S }  (  P(\bl{x} \to \bl{z}_1(\bl{x})) + P(\bl{x} \to \bl{z}_2(\bl{x})) )
\eqa
Similar definitions hold for $f_{ex,new}^I$ and $f_{ex,new}^{II}$. 
In some cases, when a point in ${\cal{X}}_b^i$ is transmitted, there may be no points belonging to ${\cal{X}}_{1-b}^i$ on one side. For instance, suppose some point in ${\cal{X}}_0^1$ such as $0001$ is transmitted. 
Then, all points in ${\cal{X}}_1^1$ are to one side of the transmitted point. 
In such a case, $\bl{z}_1(.)$ is set in the usual manner by choosing from these points, and $\bl{z}_2(.)$ is set to a special symbol $\aleph$, with the understanding that the PEP $P(\bl{x}\rightarrow \aleph)=0$ for all $\bl{x}$ in the constellation. 
Here, $\aleph$ has the interpretation of a point in the extended constellation at infinite distance from the regular points of the constellation.

The above methods are now extended to the case $d>1$ as follows. 
Suppose $\bul{x} \in {\cal{X}}_{\bul{c}}^{\bul{S}}$ where $\bul{c} = (c_1,\hdots,c_d)$, $\bul{S} = (i_1,\hdots,i_d)$ and $\bul{x} = (\bl{x}_1,\hdots,\bl{x}_d)$. 
Each $\bl{x}_l \in {\cal{X}}_{c_l}^{i_l}$ for $1\leq l \leq d$. 
Let the set $Z_{\bul{x}}$ consist of  sequences of length $d$, where the $l^{\mbox{th}}$ element is either $\bl{z}_1(\bl{x}_l)$ or $\bl{z}_2(\bl{x}_l)$. Clearly, $|Z_{\bul{x}}| = 2^d$. 
For any two signal sequences, $\bul{x}$ and $\bul{z}$, the PEP $P(\bul{x} \rightarrow \bul{z})$ is set to zero if any element in  $\bul{z}$ is $\aleph$. 
Define
\bq
f_{ex,new}(d,\mu,{\cal{X}}) = \frac 1 {m^{d} 2^{md}} \sum_{\bul{S},\bul{U}} \sum_{\bul{x} \in {\cal{X}}_{\bul{c}}^{\bul{S}}} \sum_{\bul{z} \in Z_{\bul{x}}} P\left( \bul{x} \rightarrow \bul{z}\right) \label{eq:fexNew}
\eq
Since ${\cal{X}}_{\bul{c}}^{\bul{S}}$ is a product set, 
the union bound arguments developed for the case when $d=1$ readily extend to the case when $d>1$ to yield the following upper bound.
\bq
f(d,\mu,{\cal{X}}) \leq f_{ex,new}(d,\mu,{\cal{X}})
\eq
A computationally efficient form of $f_{ex}$ is derived in \cite[(48)]{Caire98}. 
The revised expurgated bound $f_{ex,new}$ can be expressed in a similar form, with $\psi_{ex}(s)$ replaced by $\psi_{ex,new}(s)$ 
given by
\[
\frac 1 {m2^m} \sum_{i=1}^m \sum_{b=0}^1 \sum_{ \bl{x} \in {\cal{X}}_b^i} \left\{ \phi_{ \Delta \left( \bl{x}, \bl{z}_1(\bl{x}) \right)}(s) + \phi_{\Delta\left( \bl{x}, \bl{z}_2(\bl{x}) \right)}(s) \right\}
\]
The asymptotic behavior of $f_{ex}$ at large SNR ($\sigma << 1$) in the presence of fading is given in \cite[(62)]{Caire98}. 
A similar expression can be derived for $f_{ex,new}$ with $d_h$ replaced by $d_{h_c}$, where $d_{h_c}^{-2}$ is given by 
\[\frac 1 {m2^m} \sum_{i=1}^m \sum_{b=0}^1 \sum_{\bl{x} \in {\cal{X}}_b^i} \left\{ \frac 1 {|\bl{x}-\bl{z}_1(\bl{x})|^2} + \frac 1 {|\bl{x}-\bl{z}_2(\bl{x})|^2} \right\}
\]
In the design guidelines listed in \cite[\S V]{Caire98},  the harmonic mean square distance $d_h^2$ should be substituted with $d_{h_c}^2$. 
As an illustration, Table I of \cite{Caire98} is given here with the revised harmonic mean square distances. 
Here, ${d_{h_c}^I}^2$ corresponds to $f_{ex,new}^I$, and ${d_{h_c}^{II}}^2$ corresponds to $f_{ex,new}^{II}$. 
\begin{table}[htbp]
\begin{center}
\caption{Values of $d_h^2$ for some signal sets with average energy normalized to $1$.} 
\label{table:CorrectedHarmonicDist}
\begin{tabular}{||c|c|l|l|l||} \hline \hline
${\cal{X}}$ 	& $\mu$ & $d_h^2$ & ${d_{h_c}^I}^2$ & ${d_{h_c}^{II}}^2$ \\
\hline \hline
4PSK & Gray & $2$      & $2$     & $2$\\
     & SP   & $2$      & $1.333$ & $1.333$\\
8PSK & Gray & $0.7665$ & $0.637$ & $0.750$\\
     & SP   & $0.664$  & $0.436$ & $0.468$\\
16QAM& Gray & $0.492$  & $0.457$ & $0.497$\\
     & SP   & $0.441$  & $0.261$ & $0.270$\\
64QAM& Gray & $0.144$  & $0.129$ & $0.147$\\
\hline
\end{tabular} 
\end{center}
\end{table}
\section{Numerical results} \label{sec.Conclusion}
In this section, the original expurgated bound is compared numerically with the two versions of the revised expurgated bound for a Rayleigh fading channel ($K=0$) with full CSI at the receiver. The modulation schemes considered are 16QAM and 64QAM with Gray labeling. 
The binary code used is the standard rate-$1/2$, $64$-state binary convolutional code with generators (o$133$, o$171$) used in \cite{Caire98} (also given in \cite[pp. 507]{johan99} as the (o$634$, o$564$) code). This code has a minimum distance $d_2 = 10$. 
The revised versions of the expurgated bound are numerically evaluated on the same lines as the original expurgated bound.

\begin{figure}
\begin{center}
\epsfig{file=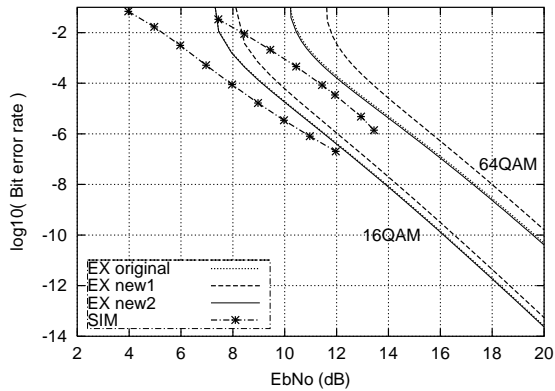,width=3in}
\caption{BER of BICM obtained from the optimal 64-state rate-1/2 code and QAM signal sets. Rayleigh fading with perfect CSI.} \label{fig:ExBounds}
\end{center}
\end{figure}

In Figure \ref{fig:ExBounds}, the bounds on BER are graphed along with simulation results, for 16QAM and 64QAM with Gray labeling. 
Curves marked by `EX orig' denote the original BICM expurgated bound, `EX new1' the bound on BER corresponding to $f_{ex,new}^I$, `EX new2' the bound corresponding to $f_{ex,new}^{II}$ and SIM computer simulation. The simulation results are obtained by using the suboptimal branch metric \cite[(9)]{Caire98}. 
The following observations can be made from the figure. 
The `EX new1' upper bound is greater than `EX orig' by a factor of about $2$ for 16QAM and about $3$ for 64QAM. 
For both 16QAM and 64QAM, the `EX new2' upper bound  is nearly indistinguishable from (but is tighter than) the original expurgated bound (EX orig) for moderate to high SNR. 
This is related to $d_{h_c}^{II}$ nearly coinciding with $d_h$ for the square QAM constellations with Gray labeling listed in Table \ref{table:CorrectedHarmonicDist}. 
While this suggests that the original expurgated bound may be a valid upper bound for square QAM with Gray labeling, we do not have a proof to support such a claim.

\bibliographystyle{ieeetr}

\begin{thebibliography}{1}

\bibitem{Caire98}
G.~Caire, G.~Taricco, and E.~Biglieri, ``Bit-interleaved coded modulation,''
  {\em IEEE Transactions on Information Theory}, vol.~44, pp.~927--946, May
  1998.

\bibitem{johan99}
R.~Johannesson and K.~Zigangirov, {\em Fundamentals of convolutional coding}.
\newblock New York : Institute of Electrical and Electronics Engineers, 1999.

\bibitem{Ungerboeck82}
G.~Ungerboeck, ``Channel coding with multilevel/phase signals,'' {\em IEEE
  Transactions on Information Theory}, vol.~28, pp.~56--67, Jan. 1982.

\bibitem{Wachs99}
U.~Wachsmann, R.~F.~H. Fischer, and J.~B. Huber, ``Multilevel codes:
  {Theoretical} concepts and practical design rules,'' {\em IEEE Transactions
  on Information Theory}, vol.~45, pp.~1361--1391, July 1999.

\bibitem{Zehavi}
{E. Zehavi}, ``8-{PSK} trellis codes for a {R}ayleigh channel,'' {\em IEEE
  Transactions on Communications}, vol.~40, pp.~873--884, May 1992.

\end{thebibliography}

\setcounter{equation}{0} 
\end{document}